\documentclass[aps,prd,preprint,superscriptaddress,showpacs]{revtex4}
\usepackage{epsfig}
\usepackage{graphicx}
\usepackage{ulem}
\usepackage{latexsym}
\usepackage{color}
\definecolor{My_red}{cmyk}{0.00,1.00,1.00,0.20}

\def\bwt{\begin{widetext}}
\def\ewt{\end{widetext}}
\def\be{\begin{equation}}
\def\ee{\end{equation}}
\def\bea{\begin{eqnarray}}
\def\eea{\end{eqnarray}}
\def\bean{\begin{eqnarray*}}
\def\eean{\end{eqnarray*}}
\def\bary{\begin{array}}
\def\eary{\end{array}}
\def\bit{\begin{itemize}}
\def\eit{\end{itemize}}

\def\nn{\nonumber}

\def\ne{\nu_e}

\begin{document}
\title{General scan in flavor parameter space in the models with vector
quark doublets and an enhancement in $B\to X_s\gamma$ process}

\author{Wenyu Wang, Zhao-Hua Xiong, Xin-Yan Zhao}
\affiliation{Institute of Theoretical Physics,
  Beijing University of Technology, Beijing 100124, China}
\date{\today}
\begin{abstract}
In the models with vector like quark doublets, the mass matrices of up and
down type quarks are related. Precise  diagonalization for the mass
matrices becames an obstacle in the numerical studies.  In this work 
we propose a diagonalization method at first. As its application,
in the standard model with one vector like quark doublet we present
quark mass spectrum, Feynman rules for the calculation of $B\to X_s\gamma$.
We find that i) under the constraints of the CKM matrix measurements, 
the mass parameters in the bilinear term are constrained to a small
value by the small deviation from unitarity; 
ii) compared with the fourth generation extension of the standard
model, there is an enhancement to $B\to X_s\gamma$ process 
in the contribution of vector like quark,
resulting  a non-decoupling effect in such models.
\end{abstract}
\maketitle

\section{Introduction}
Though the standard model (SM) has been verified to be correct times
by times, many new physics beyond standard model are proposed to
solve both experimental and aesthetical problems,  such as neutrino
masses, $\mu$ anomalous magnetic movement problem or hierarchy
problem, etc.  Many new models introduce  vector like particles (VLP) \cite{VLP-rev}
whose right handed and left handed components transform in the same
way under the weak SU(2)$\times$U(1) gauge group. The extension is acceptable
because the anomalies generated by the VLPs cancel automatically, 
and vector quarks can be heavy naturally. VLPs also arise
in some grand unification theories. For example, 
in order to explain the little hierarchy problem between
the traditional GUT scale and string scale, a testable flipped $SU(5)\times U(1)_X$
 model are proposed in Ref. \cite{Jiang:2009zza}
in which the TeV-scale VLPs were introduced~\cite{Jiang:2006hf}.
Such kind of models can be constructed from the free fermionic string
constructions at the Kac-Moody level one~\cite{Antoniadis:1988tt, Lopez:1992kg}
and from the local F-theory model~\cite{Beasley:2008dc, Jiang:2009zza}.

However when we do the flavor physics with doublet VLPs in these models~\cite{Li:2012xz, Li:2015nya},
a problem always appears when we are  dealing with the mass spectrum of quarks and leptons.
Let us start with the SM in which all fermion masses come from the
Yukawa couplings. After the spontaneously gauge symmetry breaking, we
can get two separate mass matrices $M_U,~M_D$ for the up and down
type fermions. The mass eigen states are obtained  after the
diagonalization
\begin{eqnarray}
  \label{eq:1}
Z_U^\dagger M_U U_U=M_U^D,~~ Z_D^\dagger M_D U_D=M_D^D,
\end{eqnarray}
 where
$M_U^D={\rm diag.}[m_u,m_c,m_t]$, $M_D^D={\rm diag.}[m_d,m_s,m_b]$. 
The physical measurable parameters are  $m_i$ and the so called CKM matrix
\begin{equation}
  \label{eq:defckm}
  V_{\rm CKM} = U_U^\dagger U_D.
\end{equation}
Since $M_U,~M_D$ come from separate Yukawa couplings, we can
always set one of the matrices diagonal, for example $M_U$,
and  use the CKM matrix to get the Yukawa couplings
\begin{eqnarray}
  \label{eq:11}
Z_D\left(
  \begin{array}{ccc}
    m_d & 0 & 0\\
    0 & m_s & 0\\
    0 & 0 & m_b
  \end{array}
\right)V_{\rm CKM}^\dagger=
\left(
  \begin{array}{ccc}
    Y^D_{11}v & Y^D_{12}v  & Y^D_{13}v\\
    Y^D_{21}v & Y^D_{22}v & Y^D_{23}v\\
    Y^D_{31}v & Y^D_{32}v & Y^D_{33}v
  \end{array}
\right)
\end{eqnarray}
for the calculation in flavor physics. Note that $v$ is the vacuum
expectation value (VEV) of the Higgs, and $Z_D$ is a random unitary
matrix.

Such a trick can not be used in case of the participation of a vector
doublet, namely $Q$ with gauge charge $\bf 3,~2,~\frac{1}{6}$ and
$\bar Q$ with gauge charge $\bf \bar 3,~2,~-\frac{1}{6}$, resulting
bilinear term in the lagrangian $$M^VQ\cdot\bar Q.$$ It is
clear that in the model, there are the same input parameters in the
matrices $M_U,~M_D$
\begin{eqnarray}
  \label{eq:mud}
  M_U &=& \left(  \begin{array}{cccc}
    Y^U_{11}v & Y^U_{12}v  & Y^U_{13}v & \cdots\\
    Y^U_{21}v & Y^U_{22}v & Y^U_{23}v & \cdots\\
    Y^U_{31}v & Y^U_{32}v & Y^U_{33}v & \cdots\\
    M^V_{41} & M^V_{42} & M^V_{43} &  \cdots
  \end{array}\right),~
  M_D = \left(  \begin{array}{cccc}
    Y^D_{11}v & Y^D_{12}v  & Y^D_{13}v & \cdots\\
    Y^D_{21}v & Y^D_{22}v & Y^D_{23}v & \cdots\\
    Y^D_{31}v & Y^D_{32}v & Y^D_{33}v & \cdots\\
    -M^V_{41} & -M^V_{42} & -M^V_{43} &  \cdots
  \end{array}\right).
\end{eqnarray}
The mass matrices for up and down type quarks are related to each
other. Therefore,  we can not set one of the matrices diagonal and
the CKM matrix can not be got easily. The shooting method is always
used to treat such an obstacle.  Random $M_U$ and $M_D$
are generated to meet the requirements after diagonlization: the
mass of eigen state and the measurements of elements of CKM matrix.
However this is too much time consuming, and precise solution for
diagonalization is almost unavailable.  Although this is just  a
numerical problem, when one treats the VLP contributions to
the flavor physics seriously, diagonalization of quark matrices will
be the first and important step.

In this paper, we will first propose a general method to solve the
obstacle in models with vector like quark doublets.  As its
application, we will study rare B decay $B\to X_s\gamma$ in the
SM with one vector like quark doublet. The paper is
organized as follows. We show the detail of the trick in
Section~2.  The simple application to $B\to X_s\gamma$ process,
including quark mass spectrum, Feynman rules
and the Wilson coefficients, as well as the numerical analysis for
calculation of $B\to X_s\gamma$ is shown in Section~3.
A summary is given in Section~4.

\section{The Trick of diagonalization of vector quark doublet}\label{sec2}
Firstly,  we address the problem clearly  on how to deal with the
diagonalization of $N\times N$ matrix $M_U$ and $M_D$:
\begin{eqnarray}
  \label{eq:mud2}
Z_U^\dagger M_U U_U=M_U^D ,~ Z_D^\dagger M_D U_D=M_D^D
\end{eqnarray}
in which $M_U^D, M_D^D$ are the diagonal mass matrices for up and
down type quark, respectively. Note that $N$ should be greater than 3
and the first three elements in the matices should be the three generations of quark
multiplates in the SM,  other elments with $N>3$ are the new multiplates introduced in
new physics beyond the SM. Then we have
\begin{eqnarray}
  \label{eq:mvud}
  M_U = \left(  \begin{array}{ccccc}
    Y^U_{11}v & Y^U_{12}v  & Y^U_{13}v &\cdots & M_{U1N}\\
    Y^U_{21}v & Y^U_{22}v  & Y^U_{23}v &\cdots & M_{U2N}\\
    Y^U_{31}v & Y^U_{32}v  & Y^U_{33}v &\cdots & M_{U3N}\\
    \cdots & \cdots  & \cdots  &\cdots & \cdots \\
    M^V_{N1} & M^V_{N2} & M^V_{N3}   & \cdots  & M_{UNN}
  \end{array}\right),~
  M_D = \left(  \begin{array}{ccccc}
    Y^D_{11}v & Y^D_{12}v  & Y^D_{13}v &\cdots & M_{D1N}\\
    Y^D_{21}v & Y^D_{22}v  & Y^D_{23}v &\cdots & M_{D2N}\\
    Y^D_{31}v & Y^D_{32}v  & Y^D_{33}v &\cdots & M_{D3N}\\
    \cdots & \cdots  & \cdots  &\cdots & \cdots \\
    -M^V_{N1} & -M^V_{N2} & -M^V_{N3}   & \cdots  & M_{DNN}
  \end{array}\right).
\end{eqnarray}
The last line of the two matrices has the same parameters except the
last elements.

Considering  that there are some same parameters in $M_U$ and $M_D$,
 we find that a very simple way is to add two matrices in Eq.
(\ref{eq:mvud}) 
\begin{eqnarray}
  M_U+M_D
=\left(Z_U M^D_U U_{\rm CKMN}+Z_D M^D_D \right) U^\dagger_D.
  \label{eqvud}
\end{eqnarray}
The left side of the equation is
\begin{eqnarray}
  \label{eq:pvud}
  M_U +M_D= \left(  \begin{array}{ccccc}
    Y^U_{11}v+Y^D_{11}v & Y^U_{12}v+Y^D_{12}v  & Y^U_{13}v+Y^D_{13}v &\cdots & M_{U1N}+M_{D1N}\\
    Y^U_{21}v+Y^D_{21}v & Y^U_{22}v+Y^D_{22}v  & Y^U_{23}v+Y^D_{23}v &\cdots & M_{U2N}+M_{D2N}\\
    Y^U_{31}v+Y^D_{31}v & Y^U_{32}v+Y^D_{32}v  & Y^U_{33}v+Y^D_{33}v &\cdots & M_{U3N}+M_{D3N}\\
    \cdots & \cdots  & \cdots  &\cdots & \cdots \\
    0 & 0 & 0   & \cdots  & M_{UNN}+M_{DNN}
  \end{array}\right).
\end{eqnarray}
Obviously, the mass inputs from bilinear terms vanish. We can denote the matrix in the form as
\begin{eqnarray}
M_U+M_D =M_{UD}=  \left(  \begin{array}{cc}
    {\bf M_{A}} & {\bf M_{B}}  \\
    {\bf M_{0}}& M_C
  \end{array}\right),
\end{eqnarray}
in which ${\bf M_{A}}$, ${\bf M_{B}}$, ${\bf M_{0}}$ are  $(N-1)\times(N-1)$,
 $(N-1)\times1$ and  $1\times(N-1)$  matrices correspondingly.

To prepare for the diagonalization, we chose the diagonal mass matrix elements of
quarks $(m_u,m_c,m_t,\cdots m_X)$, $(m_d,m_s,m_b,\cdots, m_Y)$
and a matrix $U_{\rm CKMN}$, which are determined partly by
experimental measurements as input parameters
\begin{eqnarray}
  \label{eq:ukmall}
  U_{\rm CKMN} &=& U^\dagger_U U_D  =  \left(  \begin{array}{cc}
    \left(U_{\rm CKM}\right)_{3\times 3}  &   \cdots  \\
  \cdots   & U_{NN}
  \end{array}\right)
=  \left(  \begin{array}{cc}
    \left(  \begin{array}{ccc}
    U_{ud} & U_{us} & U_{ub}  \\
    U_{cd} & U_{cs} & U_{cb}  \\
    U_{td} & U_{ts} & U_{tb}
  \end{array}\right)  &   \cdots  \\
  \cdots   & U_{NN}
  \end{array}\right).
  \label{CKM4}
\end{eqnarray}
Note that above $Z_U,~Z_D,~U_U,~U_D$ are unitary matrices,
but $\left(U_{\rm CKM}\right)_{3\times 3}$ is not an ordinary CKM
matrix $V_{\rm CKM}$ which is non-unitary in this case. Detailed dicussion
will be shown in the following section.

What we need to do for the next is to generate a unitary matrix $U_D$. In
the similar way we denote $U_D$ as
\begin{eqnarray}
U_D =  \left(  \begin{array}{cc}
    {\bf U_{DA}} & {\bf U_{DB}}  \\
    {\bf U_{D0}}& U_{DNN}
  \end{array}\right).
\end{eqnarray}
Both sides of Eq. (\ref{eqvud}) times the matrix $U_D$, we can get
\begin{eqnarray}
M_{UD}U_D &=&  \left(  \begin{array}{cc}
    {\bf M_{A}}{\bf U_{DA}}+{\bf M_{B}}{\bf U_{D0}} & {\bf M_{A}}{\bf U_{DB}} + {\bf M_{B}}U_{DNN} \\
   M_C{\bf U_{D0}}  & M_C U_{DNN}
  \end{array}\right)\nn\\
&=& \left(Z_U M^D_U U_{\rm CKMN}+Z_D M^D_D \right).
\end{eqnarray}
From above equation, we can get the last line of $U_D$ simply by
inputting $M^D_U,~M^D_D, ~U_{\rm CKMN}$ and random $Z_U$, $Z_D$:
\begin{eqnarray}
\left(Z_U M^D_U U_{\rm CKMN}+Z_D M^D_D \right)_{\mbox{last line}}&=&
 \left(  \begin{array}{cc}    M_C{\bf U_{D0}}  & M_C U_{DNN}
  \end{array}\right)\nn\\
&=&  M_C {\bf U_{D}}_N  ,
\end{eqnarray}
where
\begin{eqnarray}
{\bf U_{D}}_N&=& \left(  \begin{array}{cccc}  U_{DN1} &  U_{DN2}& \cdots  & U_{DNN}
  \end{array}\right)
\end{eqnarray} is a unit vector in  $N$ dimension.

Next we use the unit vector to generate total $U_D$. Since 
${\bf M_{A}}$  and ${\bf M_{B}}$ are random matrix, $U_D$ can be random too.
The unit vector  ${\bf U_{D}}_{N-1}$ of $U_D$
can be determined as
\begin{eqnarray}
{\bf U_{D}}_{N-1}&=& \left(
\begin{array}{ccccc}  -\frac{U_{DN2}^\ast}{\sqrt{|U_{DN1}|^2+|U_{DN2}|^2}} &
\frac{U_{DN1}^\ast}{\sqrt{|U_{DN1}|^2+|U_{DN2}|^2}} & 0 &\cdots
& 0
\end{array}\right).
\end{eqnarray}
It is clear that the vector is orthogonal to ${\bf U_{D}}_N$ and
normalized to 1. Then we use the first three elements of ${\bf
U_{D}}_N$ and ${\bf U_{D}}_{N-1}$ to generate ${\bf U_{D}}_{N-2}$:
Normalize the algebraic complements of first line of the $3\times 3$
matrix. Step by step, we can finally get (${\bf U_{D}}_1$, ${\bf
U_{D}}_2$, $\cdots$, ${\bf U_{D}}_{N-1}$) and form a
special $U_D^S$
\begin{eqnarray}
  \label{eq:mvud2}
  U_D^S &=&\left(
    \begin{array}{c}
      {\bf U_{D}}_1\\
      \cdots\\
      {\bf U_{D}}_{N-2}\\
      {\bf U_{D}}_{N-1}\\
      {\bf U_{D}}_N
    \end{array}
\right) =\left(  \begin{array}{ccccc}
    U_{D11}   & U_{D12}    & U_{D13}   &\cdots & U_{D1N}  \\
    \cdots & \cdots  & \cdots &\cdots & 0\\
    U_{D(N-2)1} & U_{D(N-2)2} & U_{D(N-2)3}   &\cdots & 0\\
    U_{D(N-1)1} & U_{D(N-1)2} & 0   & \cdots  & 0 \\
    U_{DN1} & U_{DN2} & U_{DN3}   & \cdots  & U_{DNN}
  \end{array}\right).
\end{eqnarray}

From above steps, we can see that (${\bf U_{D}}_1$, ${\bf U_{D}}_2$,
$\cdots$, ${\bf U_{D}}_{N-1}$) can be rotated into  any other
orthogonal $N-1$ vectors to construct random matrix ${\bf M_{A}}$
and ${\bf M_{B}}$,  only ${\bf U_{D}}_N$ must be kept unchanged.
Therefore,  a general unitary matrix can be realized by timesing a
unitary $N\times N$ matrix $U_R$,
\begin{eqnarray}
  U_D = U_R U^S_D =\left(\begin{array}{cc}
    {\bf U_R}_{N-1}   & \bf 0 \\
    \bf 0 & 1
  \end{array}\right) U^S_D
\end{eqnarray}
in which ${\bf U_R}_{N-1}$ is a $(N-1)\times (N-1)$ unitary matrix. We finish
the work by
\begin{eqnarray}
  U^\dagger_U &=& U_{\rm CKMN}U^\dagger_D\\
  M_U  &=& Z_U M^D_U U^\dagger_U\\
  M_D  &=& Z_D M^D_D U^\dagger_D
 \end{eqnarray}

At this stage, we would like to summarize our method here
\begin{itemize}
\item Step 1: Chose ($m_u,m_c,m_t,\cdots, m_X,m_d,m_s,m_b,\cdots, m_Y$) and $U_{\rm CKMN}$
 and  generate random unitary matrices $Z_U$ and $Z_D$  as the inputs for the model;

\item Step 2:  Determine the last line of matrix $Z_U M^D_U U_{\rm CKMN}+Z_D M^D_D$ as
\begin{eqnarray}
 M_C\left(  \begin{array}{cccc}  U_{DN1} &  U_{DN2}& \cdots  & U_{DNN}
  \end{array}\right)
\end{eqnarray}
and normalize it into  a unit vector ${\bf U_D}_N$.
\item Step 3: Use the unit vector ${\bf U_D}_N$ to generate other $N-1$ unitary vectors
(${\bf U_{D}}_1$, ${\bf U_{D}}_2$, $\cdots$, ${\bf U_{D}}_{N-1}$),
and form a special $U^S_D$
\begin{eqnarray}
  \label{eq:mvud3}
  U_D^S &=&\left( 
    \begin{array}{ccccc}
      {\bf U_{D}}_1 &  \cdots &  {\bf U_{D}}_{N-2} & {\bf U_{D}}_{N-1} & {\bf U_{D}}_N
    \end{array}
\right)^T.
\end{eqnarray}
\item Step 4: Generate a $N-1$ unitary matrix ${\bf U_R}_{N-1}$ to form a unitary matrix $U_R$
which is
\begin{eqnarray}
  U_R =\left(\begin{array}{cc}
     {\bf U_R}_{N-1}  & \bf 0 \\
    \bf 0 & 1
  \end{array}\right),
\end{eqnarray}
then, a general $U_D$ is obtained  by
 \begin{eqnarray}
  U_D = U_R U^S_D.
\end{eqnarray}
\item Step 5: Use these equations
\begin{eqnarray}
  U^\dagger_U &=& U_{\rm CKMN}U^\dagger_D,\nonumber\\
  M_U  &=& Z_U M^D_U U^\dagger_U,\nonumber\\
  M_D  &=& Z_D M^D_D U^\dagger_D,
 \end{eqnarray}
to get the inputs for the flavor physics.
\end{itemize}

We can see that by this trick we can skip the inputs of the bilinear
mass terms $M^V_{Ni}$. In physical analysis, the mass of eigen states
$m_{X,~Y}$ in the VLP models are inputs freely. $Z_{U}$
and $Z_{D}$ can be generated randomly, $U_{U}$ and $U_D$ can also be
scanned the most generally if we vary $U_R$ randomly. Thus the
method can do the most general scan in the parameter space of mass
matrices in the models with VLPs for the numerical studies,  which
will be shown in the following section.

\section{$B\to X_s\gamma$ process in extension of the SM with one vector like quark doublet}
\label{sec3}
\subsection{The standard  model with vector like quarks}
\begin{table}
  \centering
\caption{A simple extension of the standard  model with one vector
like quarks doublet}\label{tab1}
  \begin{tabular}{cc}
  \begin{tabular}{c|c}
      & $\bf SU(3),~SU(2),~U(1)$\\
\hline
$ Q=\left(\begin{array}{c}
 U\\
 D\end{array}\right)_L$ & $\bf 3,~2,~\frac{1}{6}$\\
 $u_{R}$  &  $\bf 3,~1,~\frac{2}{3}$ \\
 $d_{R}$  &  $\bf 3,~1,~-\frac{1}{3}$
  \end{tabular}
 &   \begin{tabular}{c|c}
      & $\bf SU(3),~SU(2),~U(1)$\\
\hline
$ V_{Q}=\left(\begin{array}{c}
\bar{V}_{d}\\
\bar{V}_{u}\end{array}\right)_{R}$ & $\bf \bar{3},~2,~-\frac{1}{6}$\\
 $\bar{V}_{uL}$  &  $\bf \bar{3},~1,~-\frac{2}{3}$ \\
 $\bar{V}_{dL}$  &  $\bf \bar{3},~1,~\frac{1}{3}$
  \end{tabular}
  \end{tabular}
\end{table}
As an application of  the method,  in this section we study the VLP
contribution to $B\to X_s\gamma$  in  a very simple VLP extension of
SM for the demonstration. In the Tab. \ref{tab1}, we list the gauge
symmetry of the matter multiplates in which the first two queues
show the quarks in the SM and the last two queues show the VLPs
with the anti-gauge symmetry. Note that we ignore partners of the last two queues
whose gauge symmetry are exactly the same as the first two queues of the SM.
As talked in the introduction, these VLPs can be heavy naturally.
Since gauge symmetry of Higgs $H=(h^+,~h^0)^{\rm T}$ is ($\bf
1,~2,~1/2$), the lagrangian for two quarks  of the model is written as:
\begin{eqnarray}
\mathcal{L}=&&Y_{d}\bar{Q}Hd_{R}+Y_{u}\bar{Q}\cdot\bar{H}u_{R}+
Y_{Vu}\bar{V_{Q}}H\bar{V}_{uL}+Y_{Vd}\bar{V_{Q}}\cdot\bar{H}\bar{V}_{dL}\nn\\
    &&+M_{Q}V_{q}\cdot Q+M_{u}\bar{V}_{uL}u_{R}+M_{d}\bar{V}_{dL}d_{R}+h.c.,
\end{eqnarray}
in which $A\cdot B=\epsilon^{ij}A_i B_j$. The first line of the
lagrangian is Yukawa terms, the second line is the bilinear terms.
Note that $Y_u$, $Y_d$ are $3\times 3$ matrix, without the
bilinear terms, the model will be almost the same as the  fourth
generation standard model (SM4).

After the electro-weak symmetry breaking, we can get the mass matrices of
up and down quarks in the basis of $(u,~c,~t,~V_u)$ and $(d,~s,~b,~V_d)$:
\begin{eqnarray}
M_U= \left(\begin{array}{cccc}
Y_{u}^{11}v & Y_{u}^{12}v & Y_{u}^{13}v & M_{u}^{1}\\
Y_{u}^{21}v & Y_{u}^{22}v & Y_{u}^{23}v & M_{u}^{2}\\
Y_{u}^{31}v & Y_{u}^{32}v & Y_{u}^{33}v & M_{u}^{3}\\
-M_{Q}^{1} & -M_{Q}^{2} & -M_{Q}^{3} & Y_{V_u}v\end{array}\right),~
 M_D=\left(\begin{array}{cccc}
Y_{d}^{11}v & Y_{d}^{12}v & Y_{d}^{13}v & M_{d}^{1}\\
Y_{d}^{21}v & Y_{d}^{22}v & Y_{d}^{23}v & M_{d}^{2}\\
Y_{d}^{31}v & Y_{d}^{32}v & Y_{d}^{33}v & M_{d}^{3}\\
M_{Q}^{1} & M_{Q}^{2} & M_{Q}^{3} & Y_{V_d}v\end{array}\right),
\end{eqnarray}
where $v$ is the VEV for $H$.
The first three elements of last line of the
matrices have the same parameter, making the scan of the parameter
space very difficult. These two matrices can be diagonalized by
unitary matrices $U$ and $Z$,
\begin{eqnarray}
Z_u^\dagger M_UU_u={\rm diag.}[m_{u},m_{c},m_{t},m_{X}],\nn\\
Z_d^\dagger M_DU_d={\rm diag.}[m_{d},m_{s},m_{b},m_{Y}].
\label{massdiag}
 \end{eqnarray}
Product of the two matrices is denoted as
\begin{equation}
  \label{eq:ukm}
U_{\rm CKM4}=U_u^\dagger U_d,
\end{equation}
which is unitary $4\times 4$ matrix. We stress that the trick we introduced in the above
section seems to just give us a numerical tool for quark masses and
some quark mixing matrices, but it is important in studying
the flavor physics in such models.

For studying VLP contributions to $B\to X_s\gamma$, we now present the
Feynman rules for the interaction of  $\bar{u}_ld_j\chi^+, ~\chi=W,~G$
 and $\bar{d}_ld_jZ$ in the Feynman gauge which read:
\begin{eqnarray}
&& {\rm i}\frac{g}{\sqrt{2}}\gamma^\mu \left[g^\chi_L(i,j)P_L
+g^W_R(i,j)P_R\right],~~(\chi=W,Z),\label{gud}\\
&& {\rm i}\frac{g}{\sqrt{2}m_W}\left[g^\chi_L(i,j)P_L
+g^\chi_R(i,j)P_R\right]~ (\chi=G) \label{hud}
\end{eqnarray}
where
\begin{eqnarray}
g^W_L(i,j) &=& \sum_{m=1}^3U_u^{*mi} U_d^{m,j},\ \
  \ g^W_R(i,j) =Z_u^{*4 i}Z_d^{4j}, \label{ud_w}\\
g^G_L(i,j) &=& \sum_{k,m=1}^3
Y_{u}^{km}vZ_u^{*ki} U_d^{mj} + Y_{Vd}vZ_u^{*4i} U_d^{4j},\label{ud_yl}\\
g^G_R(i,j) &=& -\sum_{k,m=1}^3 Y_{d}^{\ast mk}vZ_d^{*k j} U_u^{mi} -
Y_{Vu}^\ast vZ_d^{*4j} U_d^{4i}. \label{ud_yr}\\
g^Z_L(i,j) &=&
-\frac{1}{\sqrt{2}\cos\theta_{W}}\left[\left(1-\frac{2}{3}\sin^{2}
\theta_{W}\right)\delta^{ij}-U_{d}^{*4i}U_{d}^{4j}\right],\label{ud_z1}\\
 g^Z_R(i,j) &=&  -\frac{1}{\sqrt{2}\cos\theta_{W}}\left[-\frac{2}{3}\sin^{2}
 \theta_{W}\delta^{ij}+Z_{d}^{*4i}Z_{d}^{4j}\right] .\label{ud_z2}
\end{eqnarray}
Note that $U(1)_{EM}$ interaction is not changed by the VLPs, thus 
the vertices of photon and quarks are still the same as those in the SM.
From above mass matrices and Feynman rules,  we can see that the  model
has two points to be explored:
\begin{itemize}
\item The CKM matrix is got from the $W^+\bar{u}_id_j$ vertex in Eq. (\ref{ud_w}) 
  \begin{equation}
    \label{eq:vkm}
  V^{ij}_{\rm CKM4} = \sum_{m=1}^3U_u^{*mi} U_d^{mj}
   =U_{\rm CKM4}^{ij}-U_u^{*4i} U_d^{4j}.
  \end{equation}
which is non-unitary for that the indexes $i,j$ range form 1 to 4,
but the summation of index $m$ is from 1 to 3. 
$V^{ij}_{\rm CKM4}$ is also a $4\times 4$ matrix of which the upper left
elements ($i,j\ne4$)  are physical measurable value of CKM matrix $V$ as in the SM.
This is the key difference between VLP models and the SM4. Nevertheless,
the loop-level flavor change neutral current (FCNC) will be changed by the Yukawa interactions, 
then the prediction of process $B\to X_s\gamma$  may be changed
significantly.
\item The last terms in Eqs.(\ref{ud_w})-(\ref{ud_z2}), which we call the ``tail terms'',
violate the gauge universality of fermions and  cause tree-level FCNC processes
induced by the processes such as  $b \to s\ell^+\ell^-$,  
then the  constraints on the parameter space need to be explored.
\end{itemize}

\subsection{Enhancement in $b\to s$ transition}
\begin{figure}[hbtp]
\begin{center}
\scalebox{0.7}{\epsfig{file=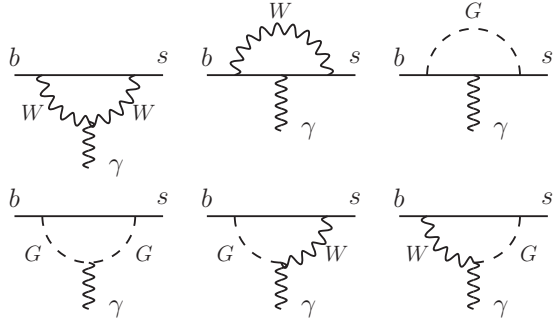}} 
\caption{Leading order Feynman diagram of $B\to X_s\gamma$ process.}
\label{fig1}
\end{center}
\end{figure}
In this subsection we focus attention on VLP contributions to
rare B decay $B\to X_s\gamma$.
The starting point for rare B decays is the determination of  the
low-energy effective Hamiltonian obtained by integrating out the
heavy degrees of freedom in the theory. For $b \to s $ transition,
this can be written as
\begin{equation}
{\cal H}_{\rm eff} = - \frac{G_F}{\sqrt{2}} V_{tb} V_{ts}^\ast
\sum_{i=1}^{10} [C_i(\mu) O_i(\mu)+C^{'}_i(\mu) O^{'}_i(\mu)]~,~\,
\label{eq:HeffBXsgamma}
\end{equation}
where the effective operators $O_i$  are same as those in the SM
defined in Ref.~\cite{BLOSM}. The chirality-flipped operators $O'_i$
are obtained from $O_i$ by the replacement $\gamma_5\to -\gamma_5$
in quark current\cite{Li:2012xz}. We calculate the Wilson
coefficient $C_7$ at matching scale $m_W$. The leading order 
Feynman diagrams are shown in FiG. \ref{fig1} and  $C_7$ reads
  \begin{eqnarray}
  C_{7}(m_W)&=&\frac{1}{V_{tb}V_{ts}^{\ast}}\sum_{i=1}^{4}\biggl[ g_{L}^{W\ast}(i,2)g_{L}^{W}(i,3)A(x_i)
 + \frac{g_{L}^{G\ast}(i,2)g_{L}^{G}(i,3)}{m_{u_i}^2} x_i B(x_i)\nn\\
& &  + \frac{g_{L}^{G\ast}(i,2)g_{R}^{G}(i,3)}{m_{u_i}m_b} x_i
C(x_i)  + \frac{g_{L}^{W\ast}(i,2)g_{R}^{G}(i,3)}{m_b}D(x_i)\nn\\
& & + \frac{m_{u_i}}{m_b}g_{L}^{W\ast}(i,2)g_{R}^{W}(i,3)E(x_i)
    + \frac{g_{L}^{G\ast}(i,2)g_{R}^{W}(i,3)}{m_b}D(x_i)\biggl] \label{fc7}
  \end{eqnarray}
where $x_i=m_{u_i}^2/m_W^2$ and the loop function
$A(x),~B(x),~C(x),~D(x),~E(x)$ are listed in the appendix.
The first two lines are the similar contribution as in the SM, while
the last lines are the terms come form the tail terms.  Note
that the contribution of right diagram in the second line of FIG. \ref{fig1}
is zero in the SM. The terms with $1/m_b$ in above equation
is extracted to compose the operator  ${\cal O}_7$.
There are two differences in the calculation of $B\to X_s\gamma$ processes
compared with the SM. One is the tail terms of gauge or Yukawa interactions, another one
is the new type of Yukawa interactions listed in Eqs. (\ref{ud_yl}, \ref{ud_yr})
which can not be written into the simple form in the SM such as
      \begin{equation}
        \label{eq:smf}
        g_L^{G,\rm SM}(i,2)=m_{u_i}V_{is},~g_R^{G,\rm SM}(i,3)=-m_{b}V_{ib}.
      \end{equation}

\begin{table}[htb]
     \caption[]{The CKM matrix elements constrained by the tree-level B decays.}
     \label{tab:c1B}
     \begin{center}
       \begin{tabular}{c||c|c}
        \hline
        & absolute value   & direct measurement from \\ \hline
        $V_{ud}$ & $0.97425 \pm 0.00022$  & nuclear beta decay \\\hline
        $V_{us}$ & $0.2252  \pm 0.0009$   & semi-leptonic K-decay\\\hline
        $V_{ub}$ & $0.00415 \pm 0.00049$  & semi-leptonic B-decay\\\hline
        $V_{cd}$ & $0.230   \pm 0.011$    & semi-leptonic D-decay\\\hline
        $V_{cs}$ & $1.006   \pm 0.023$    & (semi-)leptonic D-decay\\\hline
        $V_{cb}$ & $0.0409  \pm 0.0011$   & semi-leptonic B-decay\\\hline
        $V_{tb}$ & $0.89  \pm 0.07$       & (single) top-production\\\hline
       \end{tabular}
     \end{center}
 \end{table}

In the model with three generation quarks, the CKM matrix unitarity
is already used in the calculations of the loop-level FCNC induced
rare B decays. For consistency, in numerical analysis the
constraints on CKM matrix element are not from
processes occurred at loop level, such as rare B decays, but from
tree-level processes shown in Table~\ref{tab:c1B}~\cite{Beringer:1900zz, Eberhardt:2010bm}. 
Since there are no tree-level measurements of $V_{td}$, $V_{ts}$ now,
we use above inputs and the unitarity to get $3\times3$ unitary matrix at first.
The method is that we scan  $(V_{ud}, V_{us}, V_{ub})$ randomly
(keeping $|V_{ud}|^2+ |V_{us}|^2+ |V_{ub}|^2=1$ ) in range listed in Table~\ref{tab:c1B},
then we define two parameters  $\alpha,~\beta$ and solve them by the equations
\begin{eqnarray}
&&V_{ud}^\ast (V_{cd}+\alpha) + V_{us}^\ast(V_{cs}+\beta) + V_{ub}^\ast V_{cb} =0,\nn\\
&&\left|V_{cd}+\alpha\right|^2+  \left|V_{cs}+\beta\right|^2+  \left|V_{cb}\right|^2  =1.
\end{eqnarray}
$(V_{td}, V_{ts}, V_{tb})$ are got by the unitarity relation with 
$(V_{ud}, V_{us}, V_{ub})$ and $(V_{cd}, V_{cs}, V_{cb})$.
After that we times the $3\times3$ unitary matrix with three matrices
\begin{eqnarray}
\left(
\begin{array}{cccc}   {\bf 1} &\cdots &\cdots &\cdots \\
\cdots & \cos \theta_{4i}  & \cdots  & \sin \theta_{4i} \\
\cdots & \cdots  & {\bf 1}  & \cdots\\
\cdots & -\sin \theta_{4i}  & \cdots  & \cos \theta_{4i}
\end{array}\right)
\end{eqnarray} in which $i=1,~2,~3$ and $\max(|\theta_{41}|,~|\theta_{42}|,~|\theta_{43}|)<0.01\pi$,
to generate a $4\times4$ unitary matrix $U_{\rm CKM4}$.
$V_{\rm CKM4}$ are got by the Eq. (\ref{eq:vkm}). All the corresponding 
elements should satisfy the experiment bound list in Table~\ref{tab:c1B}
and  $V_{td}$, $V_{ts}$ ($|V_{ts}|\simeq 0.04$ which is consistent with the fitting results in 
Ref. \cite{Beringer:1900zz}) can be got too.
With this inputs in hand, the first task is to check  the scale of the
mass parameter of model, such as $M_Q$, $m_X, m_Y$. From the
$Z\bar{b}b$ vertexes in Eq. (\ref{ud_z1}) and Eq. (\ref{ud_z2}),
we can see that in order to keep
gauge universality of quarks, the tail terms in the Feynman rules
must be much smaller than the SM like terms, namely
$|Z^{4i}_{u,d}|^2_{i=1,2,3},~|U^{4i}_{u,d}|^2_{i=1,2,3}\ll
\sin^2\theta_W$. Thus in the numerical studies we require
\begin{figure}[hbtp]
\begin{center}
\scalebox{0.4}{\epsfig{file=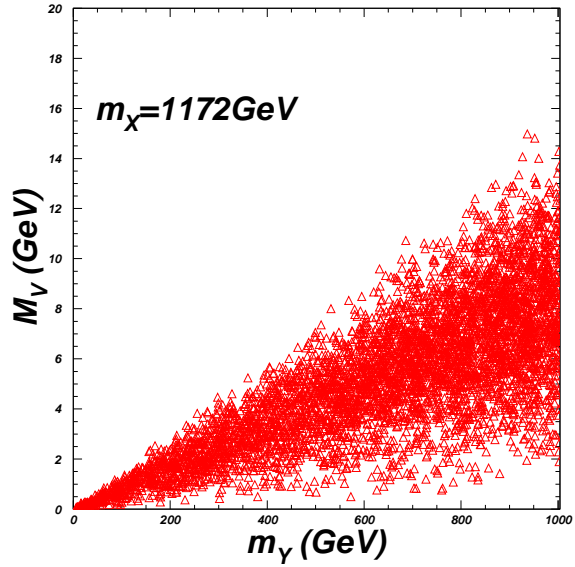}} \caption{$M_V$ versus $M_Y$ under constraints
$|Z^{4i}_{u,d}|^2_{i=1,2,3},~|U^{4i}_{u,d}|^2_{i=1,2,3}<
10^{-4}.$}\label{fig2}
\end{center}
\end{figure}
\begin{equation}
  \label{eq:vud2}
  |Z^{4i}_{u,d}|^2_{i=1,2,3},~|U^{4i}_{u,d}|^2_{i=1,2,3}< 10^{-4}.
\end{equation}
Note that though these elements are greater than $\lambda^3$ 
(parameter in the Wolfenstein parameterization \cite{Wolfenstein:1983yz}),
they are much smaller than the product of $V_{\rm CKM3}^\dagger V_{\rm CKM3}$
(almost equals $\bf 1$), thus the requirements are suitable for indicating
the contraints from the deviation from unitarity.

Since the scanning in the parameter space is freely, we set $ m_X = 1172{\rm
GeV}$ (mass of top quark plus 1000 GeV) and  scan $m_Y$ in the range of $(4.2, 1004) \rm GeV$
(mass of bottom quark plus 1000 GeV),
and $Z_{u,d},~U_{u,d}$ randomly (ignoring the CP phases).  $M_V$ is defined by
\begin{equation}
  \label{eq:mv}
  M_V = \max(|M_Q^1|,~|M_Q^2|,~|M_Q^3|).
\end{equation} 
The result for $M_V$ versus $M_Y$ is shown in the FIG. \ref{fig2}
which checks the mass input of vector doublet.  We can see that $M_V$
increases as  $m_Y$ growing up. However $M_V$ is much smaller than
$m_X$ and $m_Y$. Small mixings lead to  parameter $M_Q$ which
determine the mixing between SM quarks and vector like quarks are
also suppressed.  This is in agreement with that the deviation from
unitarity is suppressed by the ratio $m/m_{X,Y}$ where $m$ denotes
generically the standard quark masses, which is  a typical result of VLP models.
\cite{Langacker:1988ur,delAguila:1982fs,delAguila:1987nn,Cheng:1991rr,Botella:2012ju}

\begin{figure}[hbtp]
\begin{center}
\scalebox{0.4}{\epsfig{file=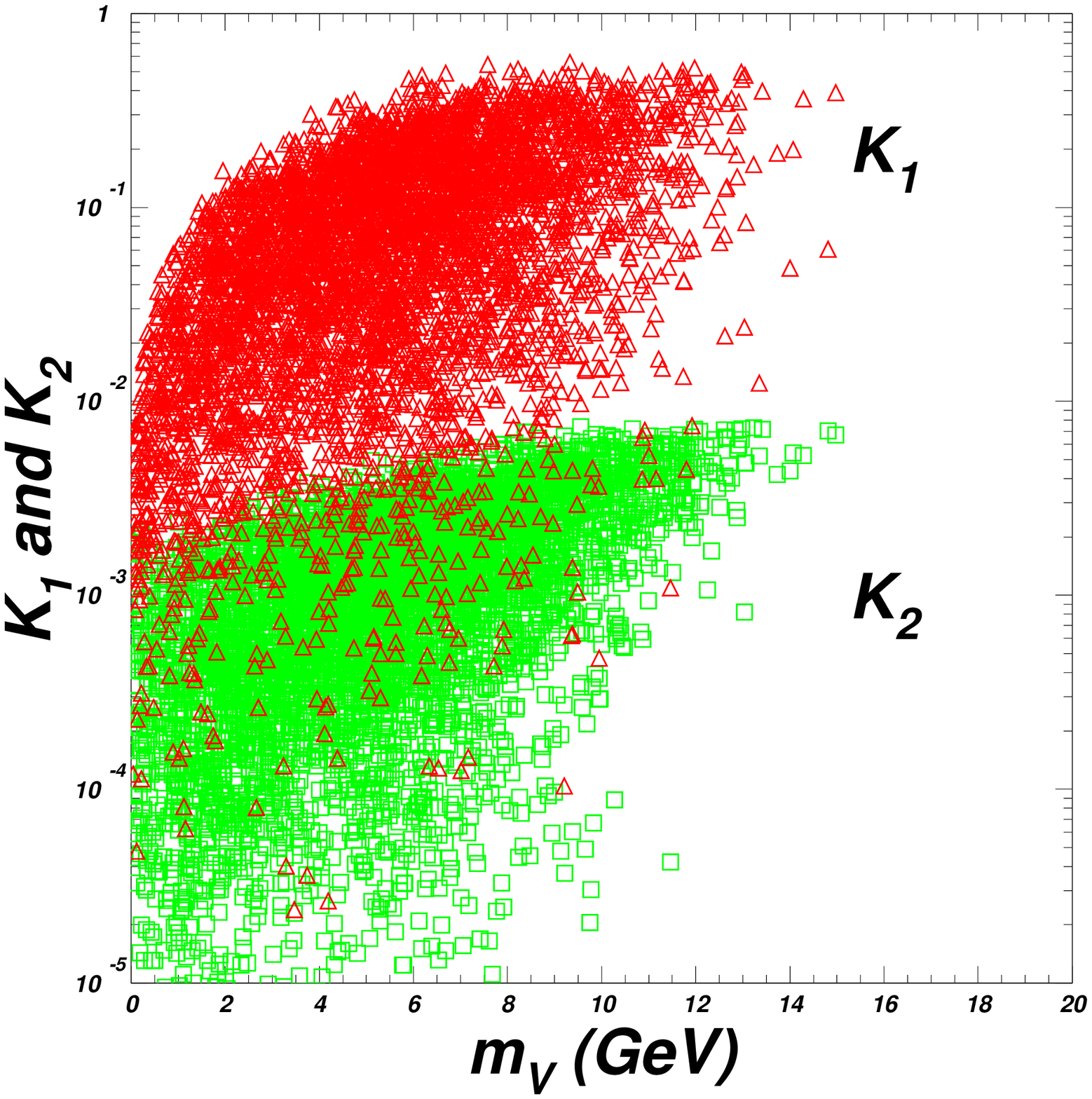}}
\scalebox{0.4}{\epsfig{file=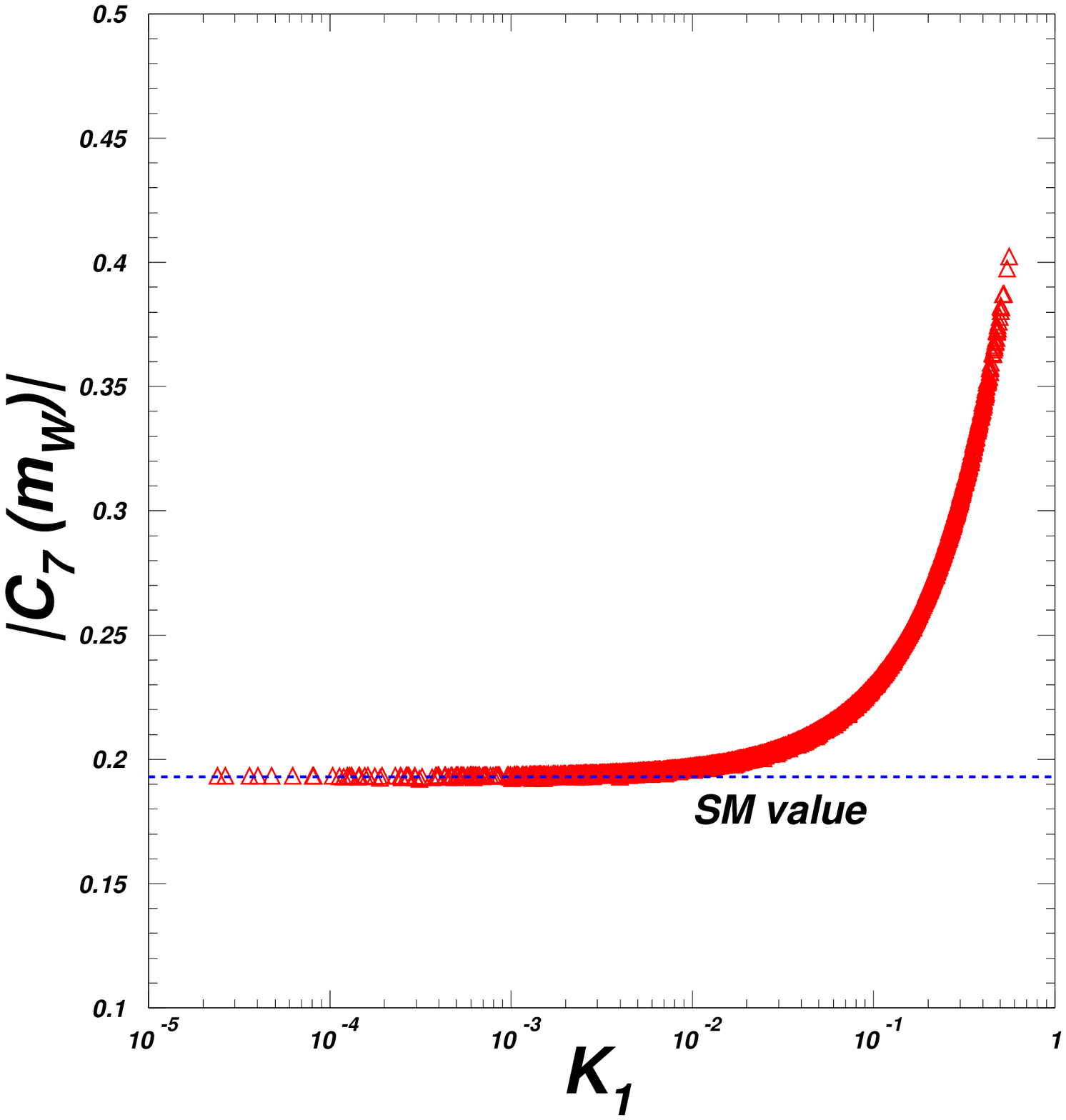}}
\caption{$K_1$, (red $\triangle$) $K_2$ (green $\Box$) versus $M_V$ and enhancement of $|C_7 (m_W)|$
case of $|Z^{4i}_{u,d}|^2_{i=1,2,3},~|U^{4i}_{u,d}|^2_{i=1,2,3}< 10^{-4}$ (color online).}\label{fig3}
\end{center}
\end{figure}
The second task  is to check the VLP contribution to $B\to X_s\gamma$. 
We find the Wilson coefficient of FCNC operator $O_7$ is
not so suppressed as the mixing.  The new contributions, from
the terms of the last line in Eq.(\ref{fc7}) are  suppressed
by the mixing, whereas the terms of first line are almost 
the same as the SM. The enhancement comes mainly from the 
$g_R^G(4,3)/m_b$ terms which are got from the
Goldstone loop in $b\to s$ transition. (The right diagram in the first line 
and diagrams in the second line of the Fig. \ref{fig1}) In order to show the
enhancement clearly, we define two factors
\begin{eqnarray}
  \label{eq:kfactor}
 K_1 &=& \frac{g_R^{G}(4,3)g_L^{G}(4,2)^\ast}{m_Xm_b V_{tb}V_{ts}^\ast}
      =  \frac{g_R^{GVb}g_L^{GVs\ast}}{m_Xm_b V_{tb}V_{ts}^\ast}\,, \\
 K_2 &=& \frac{U^{43}U^{42\ast}}{V_{tb}V_{ts}^\ast}
      =  \frac{U^{Vb}U^{Vs\ast}}{V_{tb}V_{ts}^\ast}\,,
\end{eqnarray}
in which $K_2$ denotes the deviation from the unitarity of $3\times 3$ CKM matrix,
while $K_1$ shows the enhancement of the contribution from vector like particles.
$K_1$ is in fact got from the coefficient of first term in the second line of analytical
expression of $C_7(m_W)$ in Eq. (\ref{fc7}) when $i=4$. It will be changed into exactly 
$K_2$ in case of the SM4. Note that other terms with $g_R^G(4,3)/m_b$ can 
give enhancement too, we chose factor $K_1$ for a typical demonstration  since
it seems that it will be suppressed by $m_X$. Results are shown in the
FIG \ref{fig3} in which the left panel shows $K_1$ and $K_2$ versus
$M_V$ while the right panel shows $|C_7(m_W)|$ versus $K_1$. From the left
panel, we can see that though $K_2$ increase as $M_V$ increases,  it
is still much smaller than $V_{tb}V_{ts}^\ast$, implying that
deviation of unitarity are negligible. However the factor $K_1$ can
be enhanced up to order ${\cal O}(1)$  by the increase of $M_V$. 
From the right panel, we can see that $K_1$ enhances $C_7$ up to a 
value much larger that the result of the SM. The reason for the 
enhancement mainly comes from the new type of Yukawa couplings. 
Combining of Eqs. (\ref{massdiag},~\ref{ud_yl},~\ref{ud_yr}), one
can get similar form compared with the SM4
\begin{eqnarray}
 \label{eq:comsm}
  \frac{g^G_L(4,2)}{m_{X}} &=& U_{CKM4}^{42} \label{eqglvs}
              +\frac{1}{m_{X}}\left[\sum_{m=1}^3\left(M_{Q}^{m}U_{d}^{m2}Z_{u}^{\ast44}
-M_{u}^{m}U_{d}^{42}Z_{u}^{\ast m4}\right)
+(Y_{Vd}-Y_{Vu})U_{d}^{42}Z_{u}^{\ast44}v\right],\\
\frac{g^G_R(4,3)}{m_{b}} &=&-U_{CKM4}^{\ast 43} \label{eqgrvb}
            -\frac{1}{m_{b}} \left[\sum_{m=1}^3
\left(M_{Q}^{\ast m}U_{u}^{\ast m4}Z_{d}^{43}+M_d^{\ast m}U_{u}^{\ast44}Z_{d}^{m3}\right)
+(Y^*_{Vd}-Y_{Vu}^\ast)U_{u}^{\ast44}Z_{d}^{43}v\right].
\end{eqnarray}
Since $m_X\simeq Y_{Vu}v$, $Z_u^{44},~U_u^{44}\simeq 1$, one can easily obtain that
\begin{equation}
  \label{eq:enfac1}
  \frac{g^G_L(4,2)}{m_{X}}\sim  V_{\rm CKM4}^{42},
\end{equation}
the suppression of $Z_{d}^{43}$ (order of $m/m_{X,Y}$) in Eq. (\ref{eqgrvb}) are enhanced by
terms with factor such as $\frac{Y_{Vu}v}{m_b}$, etc., resulting
\begin{equation}
  \label{eq:ehnfac2}
  \frac{g^G_R(4,3)}{m_{b}}\gg V_{\rm CKM4}^{43}.
\end{equation}
Thus the term $V_{4b}V_{4s}^\ast$ satisfying  the unitary constraint
\begin{equation}
  \label{eq:enh4}
  V_{ub}V_{us}^\ast+V_{cb}V_{cs}^\ast+V_{tb}V_{ts}^\ast
  +V_{4b}V_{4s}^\ast=0
\end{equation}
is enhanced greatly by heavy VLPs, then the factor leads the
enhancement to $C_7$. This is different from those in the SM4 
in which the contribution from the fourth generation can be neglected.

\begin{figure}[hbtp]
\begin{center}
\scalebox{0.4}{\epsfig{file=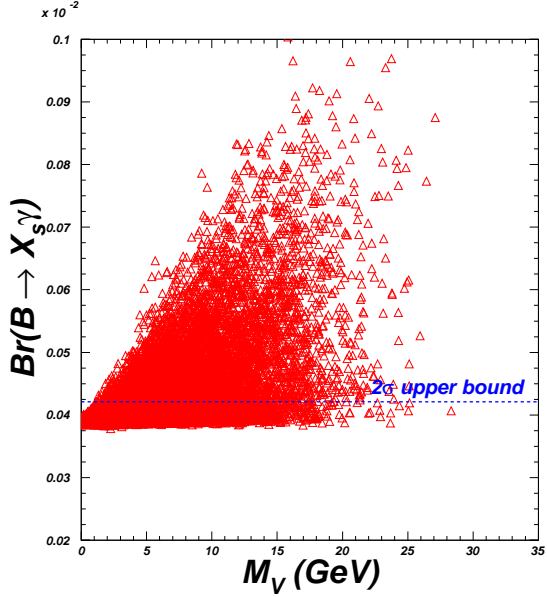}}
\caption{$B\to X_s \gamma$ prediction in random scan.}\label{fig4}
\end{center}
\end{figure}
In the numerical scan, we vary $Z_{u,d}$ and $U_{u,d}$ randomly, keeping the constraints of
$|V^{4i}_{u,d}|^2_{i=1,2,3},~|U^{4i}_{u,d}|^2_{i=1,2,3}$, 
scan $m_X$ and $m_Y$ in the range of  $(1, 2000) \rm GeV$.
Apart from the CKM limits, we use the $B\to X_s \gamma$ process 
to constrain parameter space. The branching ratio of $B\to X_s\gamma$ 
is normalized by the process $B\rightarrow X_{c}e\bar{\nu_{e}}$: 
 \begin{equation}
  {\rm Br}(B\rightarrow X_{s}\gamma)={\rm Br}^{\rm ex} (B\rightarrow X_{c}e\bar{\nu_{e}})
  \frac{|V_{ts}^{\ast}V_{tb}|^{2}}{|V_{cb}|^{2}}\frac{6\alpha}
  {\pi f(z)}[|C^{\rm eff}_{7}(\mu_b)|^{2}+|C^{\prime,\rm eff}_{7}(\mu_b)|^{2}].\label{bsg}
 \end{equation}
Here $z=\frac{m_c}{m_b}$, and $f(z)=1-8z^2+8z^6-z^8-24z^4\ln z$ is
the phase-space factor in the semi-leptonic B decay. 
The method of running of the operators from $m_W$ scale to $\mu_b$ scale can be found in
Ref. \cite{Li:2012xz}. We use the following bounds on the calculation \cite{Beringer:1900zz}
     \begin{eqnarray}
     &&   {\rm Br}^{\rm ex}(b\to ce\overline{\nu}_{e}) = (10.72 \pm 0.13) \times 10^{-2},\\
     &&   {\rm Br}^{\rm ex}(B\to X_s \gamma)= (3.55 \pm 0.24 \pm 0.09) \times 10^{-4}.
     \end{eqnarray}
The numerical results show that the $C^{\prime,\rm eff}_{7}(\mu_b)$ is much  smaller
than $C^{\rm eff}_{7}(\mu_b)$, therefore we do not present the formula of
$C^{\prime,\rm eff}_{7}(m_W)$ here.

The branching ratio as a function of $m_V$ is  shown in FIG. \ref{fig4}, 
from which we can see that ${\rm Br}({ B}\to
X_s \gamma)$ can be enhanced much greater than the experiment bound.
Then the measurements of FCNC process can give a stringent constraint on the
vector like quark model, especially when the masses of vector
quark are much greater than the electro-weak scale. 
A few remarks should be addressed:
\begin{itemize}
\item There is one point of view on the unitarity of the CKM matrix which 
is that the $3\times 3 $ ordinary quark mixing matrix is  
regarded as nearly unitary, deviation from unitarity is suppressed by 
heavy particle in the new physics beyond the SM. In other word, 
one admits that the extended CKM matrix elements exist, they
approach to zero while mass scale of the new physics approaches to infinity.
All the new  physical effects should decouple from the flavor sector and 
what should be checked is that if $3\times 3$ unitariry is consistent
in all kinds of flavor processes.
\item Another point of view is that, as in the SM case, 
the $3\times 3 $ ordinary quark mixing matrix elements are only
extracted by experiments in the measurements 
of tree and loop level precesses.  The unitarity 
should be checked, experiment measurements on the elements of matrix can be used
as the constraints to the new physics beyond the SM.
In the numerical analysis, the elements of CKM matrix
are regarded as inputs. Thus what should be done
is to scan the parameter space generally under these constraints,
no prejudice should be imposed.
Then the enhancement effect in $B\to X_s\gamma$ will be more clear. 
\end{itemize}

\begin{figure}[hbtp]
\begin{center}
\scalebox{0.4}{\epsfig{file=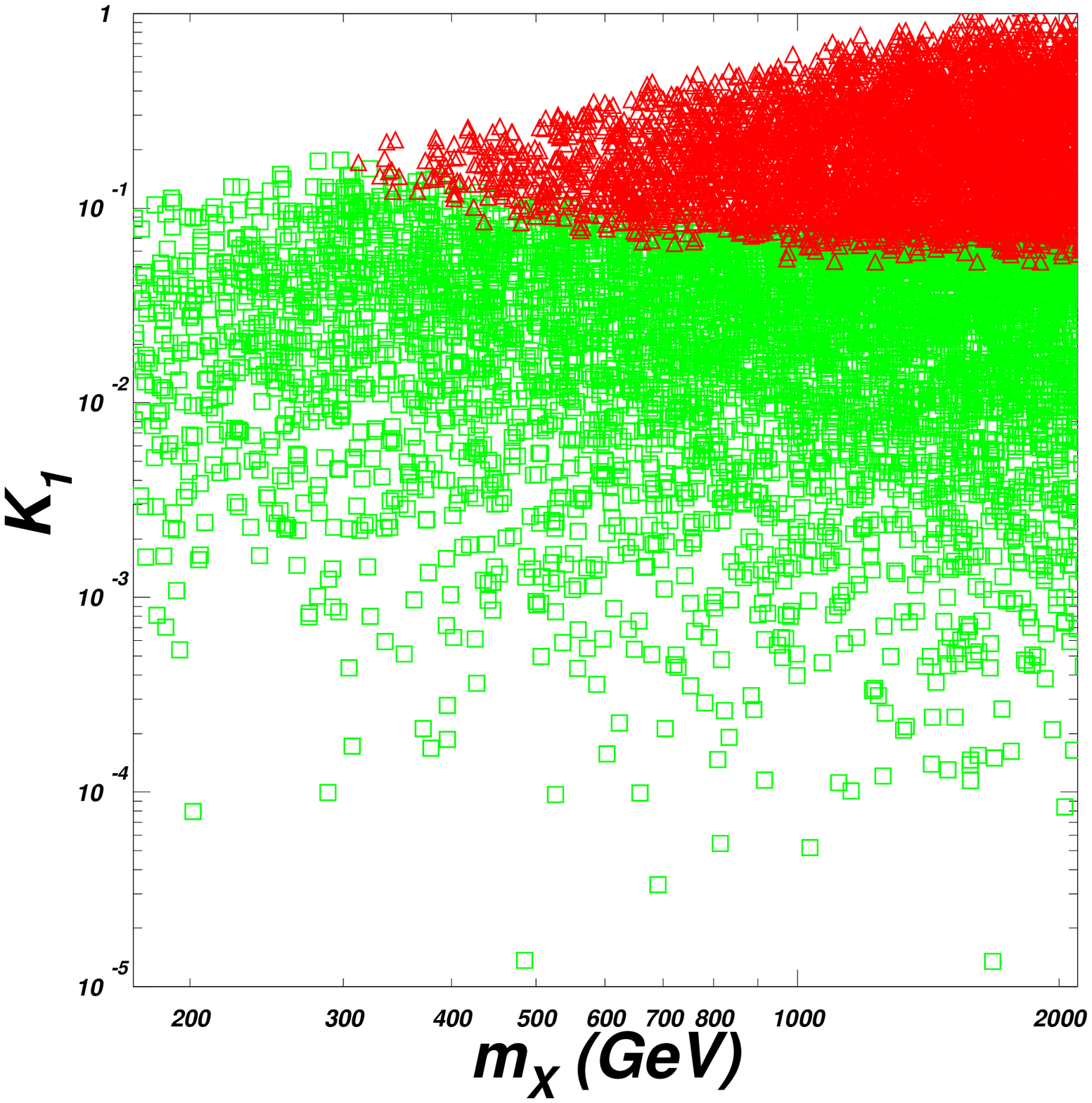}}
\scalebox{0.4}{\epsfig{file=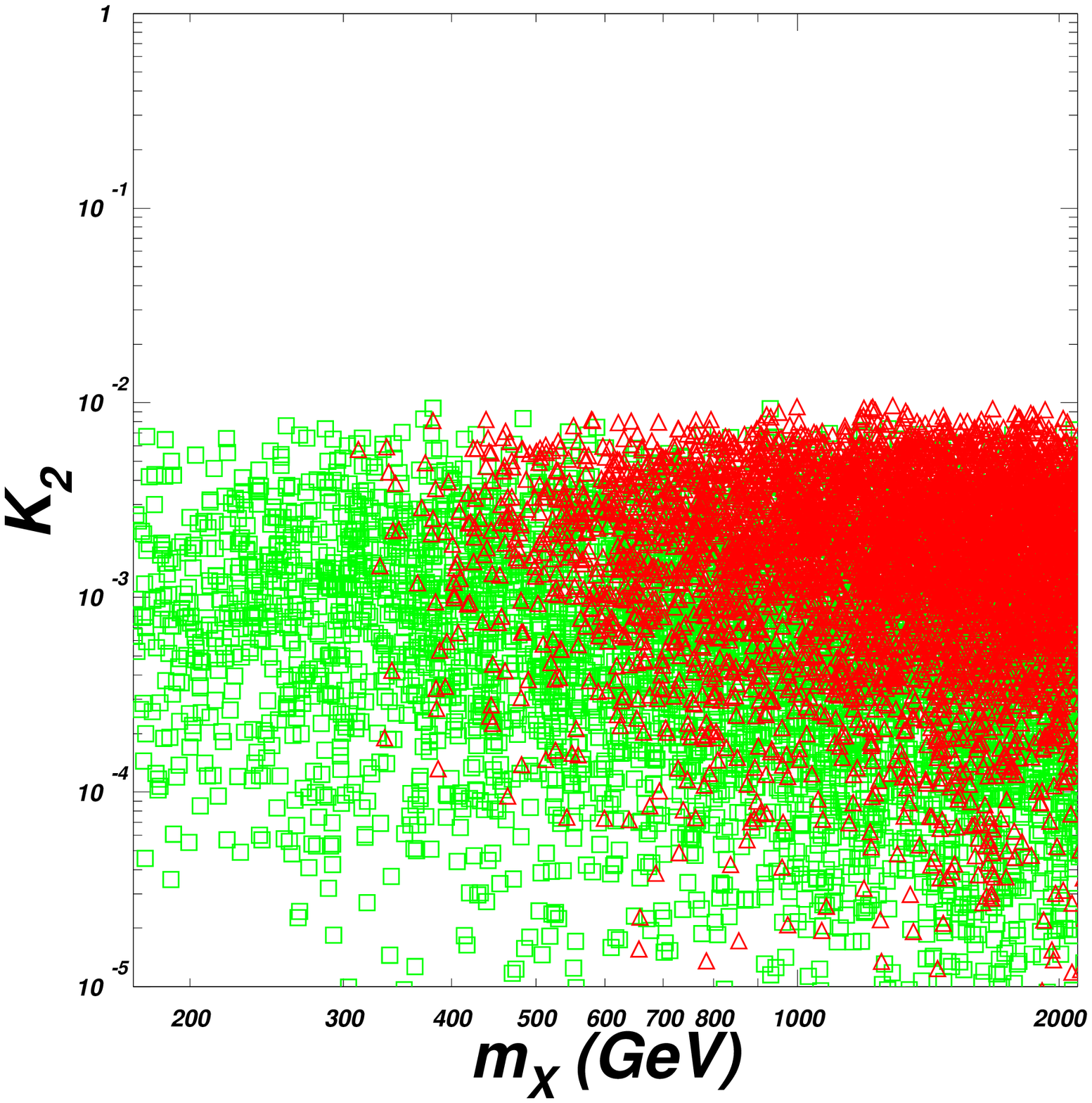}}
\caption{Enhancement factor and deviation from unitarity versus $m_X$,
red $\triangle$ are excluded by the bound of $B\to X_s \gamma$ measurement which the
green $\Box$ are the survived points (color online).}\label{fig5}
\end{center}
\end{figure}
Large parameter space is excluded by the measured branching
fraction of $B\to X_s \gamma$ as shown in FIG. \ref{fig4}.
The  enhancement effect of the VLPs can be seen in  FIG. \ref{fig5}
in which the left panel shows the enhancement factor $K_1$ 
versus $m_X$ while the right panel shows
$K_2$ versus $m_X$. From the right panel we can see that deviation
from unitarity are very small and almost irrelevant with $m_X$ since
we are doing a general scan of $Z_{u,d}$ and $U_{u,d}$. However as
we see from the left panel, as $m_X$ increases up, ${\rm
Br}({B}\to X_s \gamma)$ measurement will constrain the
enhancement factor and then constrain the input parameter of $m_X$. In
all, the enhancement can be summarized as that when mass
of vector like particle increases up, it will increase the mass
parameter $m_V$ thus give an enhancement factor under very small
deviation from  unitarity. This should be a special point when we
do the study on the vector like quark models.

\section{Summary}\label{sec4}
In the model with  vector doublets, there exist bilinear terms in 
the lagrangian, making the general scan of the Yukawa coupling very difficult.
In this paper, we show a trick to deal with the scan. Our scan method are
exactly and the more efficient. We use the trick to study a very simple extension
of the SM with  vector like quarks. We studied  one of the most important
rare B decay $B\to X_s \gamma$ process in which we found that even the deviations
from the unitarity of quark mixing matrix are small, 
the enhancement to rare B decay from VLPs are still significant. 
The enhanced effect is an important feature in the vector like particle model. 
In this work we just show the scan method, the key point of the
enhancement and how stringent constraints on the parameter space
from $B\to X_s \gamma$ measurements. What should be done includes models like extension of the SM with VLPs,
two higgs doublets models \cite{Grinstein:1990tj} or supersymmetry models \cite{Altmannshofer:2009ne}.
Such effect should be checked in all kinds of rare decays such
as inclusive process $b\to s \ell^+\ell^-$ and exclusive processes $B_s \to \mu^+\mu^-$,
$B_s \to \ell^+\ell^- \gamma$ and $B\bar B$ mixing {\it et. al.}
The detailed studies on the parameter space including other rare B decays
and new models will appear in our future work.

\begin{acknowledgments}
This work was supported by the
Natural Science Foundation of China under grant numbers 11375001
and by talents foundation of education department of Beijing.
\end{acknowledgments}
\section*{Appendix}
\begin{itemize}
\item
The loop functions for calculating the Wilson coefficients at
the matching scale are the following
\end {itemize}
\begin{eqnarray}
A(x)&=&\frac{55-170x+127x^{2}}{36(1-x)^{3}}+\frac{4x-17x^{2}+15x^{3}}{6(1-x)^{4}}\ln x,\nn\\
B(x)&=&\frac{-7+5x+8x^{2}}{36(1-x)^{3}}+\frac{-2x+3x^{2}}{6(1-x)^{4}}\ln x,\nn\\
C(x)&=& \frac{3-5x}{6(1-x)^{2}}+\frac{2-3x}{3(1-x)^{3}}\ln x,\nn\\
D(x)&=&\frac{3x-1}{4(1-x)^{2}}+\frac{x^{2}}{2(1-x)^{3}}\ln x,\nn\\
E(x)&=& \frac{-17+19x}{6(1-x)^{2}}+\frac{-8x+9x^{2}}{3(1-x)^{3}}\ln x.\nn
\end{eqnarray}

\end{document}